\documentclass{article}
\usepackage{graphicx}

\begin{document}

\title
{Quantum discrete $\phi ^4$ model at finite temperatures:\\
the mean-field phase diagram}

\author{V.V. Savkin, A.N. Rubtsov}

\date{}
\maketitle

\begin{center}
{Physics Department, Moscow State University, 119899 Moscow, Russia}
\end{center}

\begin{abstract}
{The quantum discrete $\phi ^4$ model at finite temperature is
studied in the mean-field approximation. The phase diagrams are
obtained for a wide range of the model parameters. The domains of
applicability for the classical, quantum, and order-disorder
limits are determined. It is shown that a remarkable part of the
phase diagram is reproduced by the first quantum correction to the
classical result. Parameters of Sn$_2$P$_2$S$_6$, SrTiO$_3$, and
KH$_2$PO$_4$ are placed in the phase diagram of the model.}
\end{abstract}

%\pacs{42.50.L, 77.80.B, 64.60.C}

\section{Introduction}

Quantum zero-point fluctuations play an important role in the
ferroelectric phenomena for a wide range of ferroelectric
materials \cite{Zhong}. Some of the materials, such as SrTiO$_3$,
show a displacive (soft-mode) phase transition, while others are
of the order-disorder type, for example the hydrogen-bond
KH$_2$PO$_4$. Therefore, from the fundamental point of view, one
should answer the two questions to characterize each material: (i)
what is the interplay of quantum (zero-point) and classical
(thermal) fluctuations and (ii) how close the transition is to the
order-disorder or displacive scenarios. Working with model
Hamiltonians, it is suitable to consider the quantum discrete
$\phi^4$ model here. The variation of the model parameters allows
to go continuously from the displacive to the order-disorder
second-order transition. An introduction of a finite mass and
Plank constant gives rise to the zero-point fluctuations in the
model.

Although the discrete $\phi^4$ lattice is one of the simplest
possible microscopic models for the phase-transition phenomena,
the studies have been restricted to the asymptotical cases so far.
In the general case the transition type is neither purely
displacive nor order-disorder and (or) fluctuations are neither
purely quantum nor thermal. Up to our knowledge, there is no a
quantitative investigation of this situation so far. Moreover the
exact values of the model parameters are not clear, than the known
asymptotical cases realize. This paper is aimed, in particular, to
solve the latter problem.

Let us write down the explicit formulae and discuss the known
asymptotical behaviour. The discrete $\phi^4$ model is a cubic
lattice of 2-4 anharmonic oscillators, with a nearest-neighbour
harmonic coupling. The Hamiltonian can be written as follows
\cite{Bruce}:
\begin{eqnarray} \label{H}
H&=&\sum_n \frac{P_n^2}{2 M} +V,\\  \nonumber V&=&-\frac{A}{2}
\sum_n x_n^2+\frac{B}{4}\sum_n x_n^4 +
\frac{C}{2}\sum_{nn'}(x_n-x_{n'})^2\sigma_{nn'}.
\end{eqnarray}
Here $x_n$ is the displacement of the $n$-th atom in the unit
cell, $A,B,C$ are parameters of the model. The first two terms in
the potential energy $V$ form a double-well potential and the last
term stands for the interaction. There is only the inteaction
between nearest-neighbours in the model, {\it i.e.}
$\sigma_{nn'}=1$ if $n, n'$ are the nearest neighbours and
vanishes elsewhere. Note that the Hamiltonian (1) includes the
displacement of atoms, but not the actual distance between them.

If there were no any fluctuations in the system, all atoms would
occupy the same position in the cell $x_n=\sqrt{A/B}$ (or
$x_n=-\sqrt{A/B}$). There are two sources of fluctuations. The
dimensionless mass $m=M C A^2 / \hbar^2 B^2$ determines the role
of quantum zero-point fluctuations in the model (the system is
classical at $m\to\infty$). The classical thermal fluctuations are
governed by the reduced temperature $t=T B/A C$ (the system
becomes purely quantum at $t=0$). Large fluctuations can destroy
the ordering, so that the average of $x$ vanishes. The system of a
dimensionality $d>1$ shows a second-order transition for any
positive $A, B, C$. The parameter $a=A/C$ determines the type of
the transition: displacive transition corresponds to $a \to 0$,
while the limit $a\to+\infty$ corresponds to the order-disorder
system \cite{AR,my}. At a given $a$ there is a critical line
$t_c(m_c)$ in the $t-m$ phase diagram (hereafter the subscript "c"
stands for the critical values of temperature and mass).

In the reduced dimensionless variables, the Hamiltonian takes the
form
\begin{eqnarray} \label{Va}
&&h=-\sum_n\frac {\nabla_n^2} {2m}+v, \\   \nonumber &&v=\left(2d
- \frac{a}{2}\right)\sum\limits_n {x_n^2}+{a \over 4}\sum\limits_n
{x_n^4}-\sum\limits_{n,n'} {x_n x_{n'}} \sigma_{nn'}.
\end{eqnarray}

The asymptotical cases are studied extensively. Basically, they
are the follows.

At the limit $a\to+0$ the transition is of the soft-mode type.
This asymptotic is studied at $t=0$, as well as at finite
temperature \cite{ima,ima1}. Since the double-well potential is
small for this case, it can be treated as a perturbation.
Therefore a convenient method of the investigation of this limit
is the independent-mode approach \cite{Bruce}. The system is
harmonic in the zero approximation, therefore the phonon
vibrations $Y_k\propto\sum e^{ikj}x_j$ can be treated as uncoupled
oscillators so that one assumes $<Y_k Y_l>\propto \delta(k+l)$
(hereafter the triangle brackets stand for the averaging over
thermal and quantum fluctuations). In the first approximation the
anharmonic term renormalize the phonon dispersion. Virtually, the
real interaction can be reduced to an effective ({\it i.e.}
dependent on the temperature and mass) harmonic one. At certain
values of $t$ and $m$ the zero-frequency mode (soft-mode) appears
in the system indicating the phase transition. The
independent-mode approach is exact in the $a\to+0$ limit. However,
serious difficulties appear for a finite $a$; particularly, $t_c$
and $m_c$ are incorrectly predicted to be independent on $a$ in
the independent-mode scheme.

At the opposite limit $a\to+\infty$, the discrete $\phi^4$ model
shows an order-disorder phase transition. The physical picture in
this case is the follows. Since an on-site potential is much
larger than all other energy scales, atoms are localized in either
left or in the right side of the double-well in the zero
approximation: $x_n\approx\pm\sqrt{A/B}$. Therefore, each
oscillator can be treated as a two-level system. The tunnel
coupling between the two sides of a well produces some energy
splitting $\Gamma$, which depends on the mass $m$ implicitly. This case coincides with the transverse-field
Ising model \cite{tfi0,tfi1}. This model describes an array of $\frac
{1} {2}$-spin particles in the external field with a
nearest-neighbour interaction between particles. The Hamiltonian
of this model is
\begin{equation} \label{trans}
H=-\Gamma \sum_{i}S_{i}^{x}-\frac {C} {2}\sum_{i,j} \sigma_{nn'}
S_{i}^{z}S_{j}^{z},
\end{equation}
where $S_{i}^{x}$, $S_{i}^{z}$ are $\frac {1} {2}$ spin operators.
The phase transition can be observed by changing the two
parameters: temperature $t$ and transverse field $\Gamma$. The
transverse-field Ising model is also widely studied at $t=0$ and
at finite temperature by various approximations. High-temperature
series expansions, ground-state perturbation theory,
correlated-basis-function analysis, random phase approximation and
its generalizations, $1/\Re$ and $1/S$ expansions, continuum field
theory methods have been used
\cite{tfi0,tfi1,tfi2,tfi3,tfi4,tfi5,tfi6}.

The classical case corresponds to the condition $m\to \infty$. In
this case the momentum-coordinate uncertainty is small, so that
the partition function can be factorised as $Z=\int \exp
(-\sum\limits_n p_n^2/2 m t) d^N p_n \cdot \int \exp (-v/t) d^N
x$. The temperature dependence of the order parameter is
determined here by the second term and is therefore independent of
mass $m$. The dependence $t_c(a)$ is studied both numerically and
by analytical approaches for 2D, layered and 3D systems
\cite{AR,my}. An extension of the mean-field
scheme has been developed, which is appropriate for the
description of Monte Carlo results with a good accuracy in a wide
rage of parameter $a$.

The quantum limit takes place at $t=0$ and at finite mass $m$. The
dependence $m_c(a)$ is studied \cite{AR1} by quantum Monte Carlo
simulations and the simplest analytics for a wide range of $a$.

In this paper, we study the phase diagram of the quantum discrete
$\phi^4$ model for a full range of parameters, and determine the
domain of applicability for the asymptotical results mentioned
above. We also consider the leading-order quantum correction to
the classical result. In a conclusive part of the paper the
parameters of several ferroelectric materials are placed in the
phase diagram obtained.

We chose the mean-field (MF) approximation for the study because
of its simplicity. Actually, we are interested in quite crude
characteristics such as the domains of applicability. Therefore we
believe that this method is suitable here, although it is not very
accurate. The validity of the MF approach is discussed in more
details in the end of the paper.

\section{The quantum mean-field approximation}

In the MF approach, the real interaction between particles is
replaced with an interaction of the on-site oscillator with an
external field $E=4 d <x>$. The self-consistent set of MF
equations for $<x>$ is formed by this formula and the expression
for the average displacement of the on-site oscillator \cite{LL}
\begin{equation}
    <x>=\frac{{\rm Sp}\left( x e^{-(H_0-E <x>)/T}\right)}{{\rm Sp}
    \left(e^{-(H_0-E <x>)/T}\right)},
\end{equation}
where
\begin{eqnarray} \label{H0}
    H_{0}&=&\frac{P^2}{2 M}+V_{0}, \\
    \nonumber V_{0}&=&\left(2 d C-\frac{A}{2}\right)x^2+\frac{B}{4} x^4.
\end{eqnarray}
In this paper, we are interested in the position of the transition point in phase
diagram.
The condition $<x> \to 0$ is fulfilled in the point of the
phase transition. Keeping the leading-order in $<x>$ terms, one
obtains the MF equations for the critical point:
\begin{equation} \label{mf}
4 d C \chi_{1}=1,
\end{equation}
where $\chi_1$ is the static linear susceptibility, which is
determined at the finite temperature by the standard
quantum-mechanical formula \cite{LV}:
\begin{equation} \label{chi}
    \chi_{1}=\sum_{ij} \frac{2 d_{ij}^2} {E_i-E_j}\exp{\frac{F-E_i}{T}}.
\end{equation}
Here, $F=-T\ln Z$ is the free energy ($Z$ is a partition
function), $E_i=<i|H_{0}|i>$ and $d_{ij}=<i|d|j>$ are respectively
the eigenenergies and the dipole matrix elements for the on-site
Hamiltonian $H_0$, defined in Eq. (\ref{H0}).

Let us briefly discuss the structure of the spectrum. In the
displacive limit $a\to +0$ it is almost equidistant, because the
nonlinearity of the potential is small. In the order-disorder case
of $a\to +\infty$ the situation is quite different. There are
several doublets of the lowest levels, whereas the rest of the
spectrum is situated much higher (Fig. 1). A number of doublets
increases with an increase of $m$, since the system becomes more
classical and the spectrum becomes almost continuous. At the same
time, the behaviour of the system is determined by the first
doublet at low and moderate masses. This is the case of the
transverse-field Ising model (3).

The spectrum and matrix elements of the on-site oscillator
(\ref{H0}) were calculated numerically. Here are some details of
the numerical procedure. The eigenfunctions of the Schr\"{o}dinger
equation with $H_{0}$ were found by the phase method
\cite{Kalitkin}. We controlled the accuracy of calculations by the
check of the condition $<i|j>=\delta_{ij}$, which was fulfilled
with an error bar not exceeding $10^{-3}$. We used approximately
50-70 levels of $H_{0}$ to obtain the phase diagrams $m_c(t_c)$ in
the quantum MF approach at moderate values of $a$ and for
$m\approx5$. This amount is necessary to describe correctly the
case of large mass, than the levels are quite dense. The number of
necessary levels is increased with an increase of $t$ as well.

The phase diagrams are obtained for a wide range of the parameter
$a$. They correspond to the displacive limit, order-disorder limit
and to an intermediate case. The dependencies $m_c(t_c)$ are
plotted in Fig. 2 by solid lines for the three values of the
parameter $a$ for 2D and 3D lattices. All curves show a good
agreement with the known asymptotes for the quantum limit ($t=0$)
and for the classical case ($m\to +\infty$) \cite{AR,my,AR1}.

\section{The limits}
Now we turn to the discussion of the domains of applicability for
the four mentioned asymptotical cases: displacive, order-disorder,
classical and quantum limits. In the displacive limit, the MF
scheme suffers serious difficulties (see discussion in the end of
the paper). However, from the previous studies of the classical
\cite{AR,my} and quantum \cite{AR1} systems, we can estimate that
the displacive limit occurs at $a$ smaller than $a\approx1$. For
the three other limits, we use the following scheme. We calculate
the critical parameters by the MF expressions which are valid in
the corresponding asymptotical cases, and compare them with the
direct MF calculations. For each limit we determine the areas
where the error bar does not exceed 1\% and 5\%.

For the quantum limit a deviation of $m_c$ from the value of
$m_c(t=0)$ is detected for each $a$. The two areas, where $m_c$
differs from $m_c(t=0)$ in a value smaller than 1\% and 5\%,
respectively, are marked in Fig.2.

To determine the areas, where the system can be considered as
classical, we compare $t_c(m_c)$ with the classical value
$t_c(\infty)$, at a given value of $a$. The obtained areas, where
the classical result delivers a 1\% and 5\% accuracy, are shown in
Fig.3. For comparison, we also present the "purely quantum" MF
dependence of $m_c(a)$ at $t=0$ in the same figure.

Finally, the domain of applicability of the order-disorder limit
is estimated from the comparison of the dependencies $m_c(t_c)$,
obtained from the calculation by (\ref{chi}) with a complete set
of eigenfunctions, and with only the first doublet in the
spectrum. The result is shown in Fig.3.

Thus, the values of the parameter $a$, which are used to plot the
curves in Fig.2, cover a complete range of phase transitions from
the displacive type for small values of $a$, to the
order-disorder one for large values of $a$. Besides, the presented
intervals of $m$ and $t$ (Figs.2,3) show a crossover between
quantum and classical phase transitions for the discrete $\phi ^4$
model.

\section{The quantum correction formula}
We have also plotted the leading-order quantum correction to the
classical MF value. The idea is to use usual classical MF scheme,
which takes into account quantum fluctuations in the system. It
can be obtained in Feynman's consideration of the path-integrate
representation for the partition function \cite{Feynman}. The
partition function of the on-site oscillator
$Z={\rm Sp} \left(e^{-(H_0-E <x>)/T}\right)$ in this representation can be written as
follows:
\begin{equation} \label{Feynman}
Z=\int [{\cal D} x] \exp\left(\hbar^{-1}\int_{0}^{\frac{\hbar}{T}}
L(x(\tau)) d\tau\right),
\end{equation}
where the integration is taken over all classical trajectories
$x(\tau)$, describing the propagation in the imaginary time
$\tau=0...\hbar/T$ with periodic boundary conditions applied.
$L(x(\tau))$ is a Lagrangian along the trajectory $x(\tau)$. The
semi-classical approximation consists in the assumption that the
deviations of the trajectory from its mean position $x_0=T
\hbar^{-1}\int x d \tau$ are small. One should perform a
Taylor-series expansion of ({\ref{Feynman}}) in powers of
$\tilde{x}(\tau)=x(\tau)-x_0$ and keep the terms up to
$\tilde{x}^2$. After this, $\tilde{x}$ can be integrated out,
yielding
\begin{equation} \label{corr}
Z\approx Z_0 \int d x_0 \exp-\frac{1}{T}\left( V_{0}(x_0)-E x_0
+\frac{\hbar ^2} {24MT} \frac{\partial^2 V_{0}(x_0) }{
\partial
  {x_n^2}}\right).
\end{equation}
Here $Z_0$ is a factor, independent of $V_{0}(x)$. Thus, in this
semi-classical approximation we can consider a quantum system
using its classical potential with quantum correction
\cite{Feynman}:
\begin{eqnarray} \label{qcorr}
V_q=V_{0} + \frac{\hbar^2 V_{0}''}{24 M T}.
\end{eqnarray}
Virtually, quantum correction renormalises the on-site potential
parameters making them dependent on temperature.

The results obtained from the classical MF theory with the quantum
correction ({\ref{qcorr}}) are presented in Fig.2 by the dashed
lines. All curves demonstrate an excellent agreement with ones
obtained directly from the quantum MF calculations ({\ref{mf}}).
Certainly, the approach ({\ref{qcorr}}) does not work at low $t$
and $m$ at all. In our case of discrete $\phi ^4$ model we can
calculate the phase diagrams in this way down to the mass $m
\approx 0.35$. Nevertheless, the area where the semi-classical
approximation works is large enough to demonstrate a crossover
between the quantum and classical phase transitions. We would like
also to stress that the calculation via the quantum-mechanical
formula ({\ref{mf}}) is about $~10^4-10^5$ times longer than the
classical MF method with the quantum correction ({\ref{qcorr}}).

\section{Experimental data and discussion}

The estimation of the parameters $m$ and $a$ of the discrete $\phi
^4$ model for real materials and their accordance to the limit
cases on the phase diagram $m_c(a)$ (Fig.3) is of interest. We
will discuss a situation with the following ferroelectrics:
Sn$_2$P$_2$S$_6$, SrTiO$_3$, KH$_2$PO$_4$ (KDP).

The transition in Sn$_2$P$_2$S$_6$ is of an intermediate type
\cite{vis1,Barsamian}. Therefore, there are thermal regions in
phase diagrams, where Landau theory is valid, {\it i.e.} the free
energy is of the form: $\frac{F}{V}\approx\frac {\alpha(T)
P^2}{2}+\frac{\beta(T) P^4}{4}$, with $\alpha \propto (T-T_0)$,
$\beta \approx const$. In our estimations for Sn$_2$P$_2$S$_6$, we
use the data for the Curie low and the value of the saturated
spontaneous polarization to obtain values of coefficients
$\alpha(0)$ and $\beta(0)$ \cite{Hlinka}. The following estimation
for the parameters $m$ and $a$ can be suggested (for an estimation
of $a$, our method is very similar to what was used in
\cite{Hlinka}). The following relation can be easily obtained:
$A^2/B=V_{uc} \alpha^2/\beta$ ($V_{uc}$ is a volume of the unit
cell). Also, accordingly to the phase diagram for the classical
discrete $\phi^4$ model \cite{AR}, one has $kT_c=\lambda A C/B$,
with $\lambda \approx 6$ for an intermediate type of the
transition. From these two relations one can guess about the ratio
$a=A/C\approx\lambda V_{uc} \alpha^2/k T_c \beta $. A suitable
estimation for the parameter $m=M C A^2 / \hbar^2 B^2$ comes from
the observation that the mean-square of the displacement $<\Delta
x^2>\approx A/B$ is known from the experiment \cite{vis1,Scott}.
Combining this with the two previous formulae, we obtain $m\approx
M kT_{c} <\Delta x^2> / \hbar^2 \lambda$. The values $\alpha(0)$,
$\beta(0)$, $V_{uc}$, $kT_c$ and $ <\Delta x^2>$ obtained from the
experimental data \cite{vis1,Barsamian,Hlinka,Scott}, as well as
our estimations for $m$ and $a$ are presented in Table 1.

We also use the same method in estimations for SrTiO$_3$, which is
a displacive-type material \cite{Bruce}. However, there are some
special issues in this case. An ordering in SrTiO$_3$ is absent
due to the quantum effects. The dependencies $\alpha(T)$ and
$\beta(T)$ are Landau-like only at temperature down to
approximately 50 K; at lower temperature they are saturated.
Therefore, in our estimations we use the experimental data
\cite{Christen,Grupp} for $\alpha(T), \beta(T)$ for $T=50..100K$.
We approximate these data by the Landau-like dependencies $\alpha
\propto (T-T_c)$, $\beta=const$, i.e. assuming a "classical"
picture of the phase transition. We extrapolate these "classical"
dependencies down to $T=0$, and obtain the "classical" critical
temperature, and $\alpha(0), \beta(0)$ from this extrapolation. If
the system were purely classical, $\alpha(0), \beta(0)$ and $T_c$
would take these values. The extracted data are presented in Table
1. Since the displacive phase transition takes place in SrTiO$_3$
we use the estimation for $\lambda$ as $\lambda \approx 3$,
according to the previous study of the 3D discrete $\phi ^4$ model
\cite{AR}. All other values required for estimations of $a$ and
$m$ are known from the literature \cite{Christen,Grupp,Kubo} and
are placed in Table 1 as well.

The corresponding points for Sn$_2$P$_2$S$_6$ and SrTiO$_3$ lie in
the phase diagram (Fig.3) in the classical and quantum areas,
with the intermediate and displacive types of the transitions,
respectively. These data correspond to the experimental
observations
\cite{Bruce,vis1,Barsamian,Hlinka,Scott,Christen,Grupp,Kubo}.

\begin{table}
\caption{\label{t.1}Experimental data and estimations for $m$ and
$a$ for Sn$_2$P$_2$S$_6$ and SrTiO$_3$.}
%\begin{indented}
%\item[]\begin{tabular}{@{}llllllll}
\begin{tabular}{@{}llllllll}
%\br
 &$\alpha(0)$, J$\cdot$m$\cdot$C$^{-2}$&$\beta(0)$, J$\cdot$m$^5\cdot$C$^{-4}$&$V_{uc}$, m$^3$&$kT_c$, J&$<\bigtriangleup
x^2>$, m$^2$&$m$&$a$\\
%\mr
Sn$_2$P$_2$S$_6$&$2.7\times10^9$&$10^{11}$&$2.3\times10^{-28}$&$4.6\times10^{-21}$&$10^{-21}$&$12$&$21$\\
SrTiO$_3$&$5\times10^7$&$4\times10^{9}$&$6.4\times10^{-29}$&$7\times10^{-22}$&$1.7\times10^{-22}$&$0.3$&$0.2$\\
%\br
\end{tabular}
%\end{indented}
\end{table}

However, the Landau theory is hardly applicable even qualitatively
in the case of KDP due to the order-disorder phase transition of
the essentially quantum nature (the transverse-field Ising model
is quite suitable here). We make use of the following method of
the estimation of parameters. The splitting of the first doublet
of the on-site oscillator spectrum \cite{Vaks} is known to be
about $\omega_1-\omega_0\approx 80$ K, while the third level lies
on the distance of approximately $\omega_2-\omega_0\approx 800$ K
from the first one. Further, we take into account the value for
$T_c\approx 120$ K.
 Given these data, we adjust $a$ and $m_c$ to deliver the known
 values of $(\omega_1-\omega_0)/(\omega_2-\omega_0)$ and
$(\omega_1-\omega_0)/T_c$. This estimation gives us $a\approx 250$
and $m\approx 0.1$.  Thus, the point for KDP disposes in the
quantum area of the phase diagram (Fig.3) and corresponds to an
order-disorder transition.

Thus, our estimations place the three mentioned materials to the
asymptotical cases they are commonly attributed to. Now we discuss
how the points can move on the phase diagram. The most direct way
is to change the isotope mass of the proper components. An elegant
example is the restoration of the ordering via the change from
$^{16}$O to $^{18}$O in SrTiO$_3$ \cite{SrTiO3}. The transition
temperature happens to be about 23 K in this case, while the
"classical estimation" for $T_c$ (placed in Table 1) is
approximately two times larger. This corresponds well with our
data: at small $a$ a 5-7 \% change in $m$ can increase the
transition temperature from 0 to $t_c\approx 2$, while the
"classical" asymptote is $t_c \approx 4$ (see Fig.2). Clearly,
however, the "isotopic" change in $m$ cannot be large.

Finally, we discuss the applicability of the MF method for the
present problem. As the MF scheme ignores correlation effects in
the system, it overestimates the transition temperature (or
underestimates the critical mass). This overestimation is
particularly large for the displacive limit. Namely, the MF
approach predicts a finite value of $t_c(a\to +0)$ in any
dimension, whereas actually fluctuations result in $t_c \propto
\ln^{-1} a$ in 2D at finite temperature \cite{my,toral}. However,
for the moderate and large values of $a$ in 2D, as well as for the
whole range of $a$ in 3D, we expect that the qualitative behaviour
of $t_c(a)$ and $m_c(a)$ is reproduced correctly. Quantitatively,
it is known that the MF scheme overestimates $t_c$ \cite{AR,my} or
underestimates $m_c$ \cite{AR1} by some 30-50 \%. We expect that
the similar situation takes place for the general case, then both
thermal and quantum fluctuations are present. To get more
justification, we have now started the quantum Monte Carlo
simulations for the system under consideration \cite{MC}. The
pilot calculations support the qualitative validity of the MF
results.

Authors are grateful to T.V. Murzina and T.V. Dolgova for
reading the manuscript. This work was supported by the INTAS YSF
2001/1 - 135, the RFFI foundation (grant 00-02-16253 and special
grants for young scientists) and by the "Russian Scientific
Schools" program (grant 96-1596476).

\begin{figure}[p]
  \centering
  \includegraphics[width=1.0\textwidth]{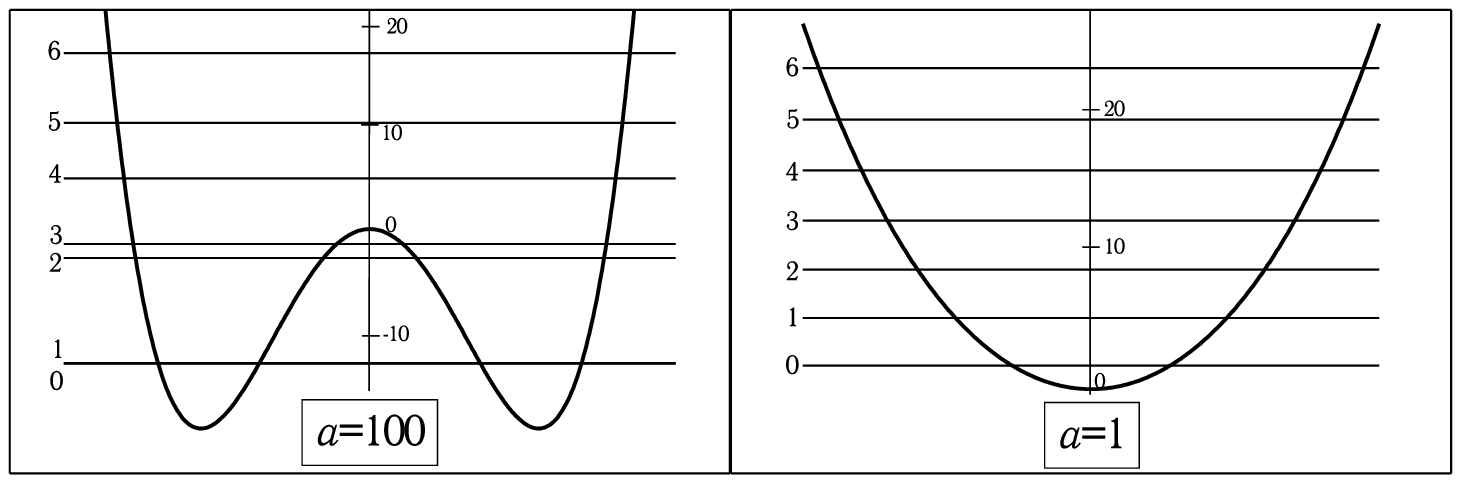}
  \caption{The on-site potential and its spectrum for the
two limits of the discrete $\phi ^4$ model. The left panel is a
typical spectrum for the order-disorder limit ($a=100$). The right
panel is a typical spectrum for the displacive limit ($a=1$).
Parameters $m=1$ and $d=3$ are used for the both cases.}
\end{figure}

\begin{figure}[p]
  \centering
  \includegraphics[width=1.0\textwidth]{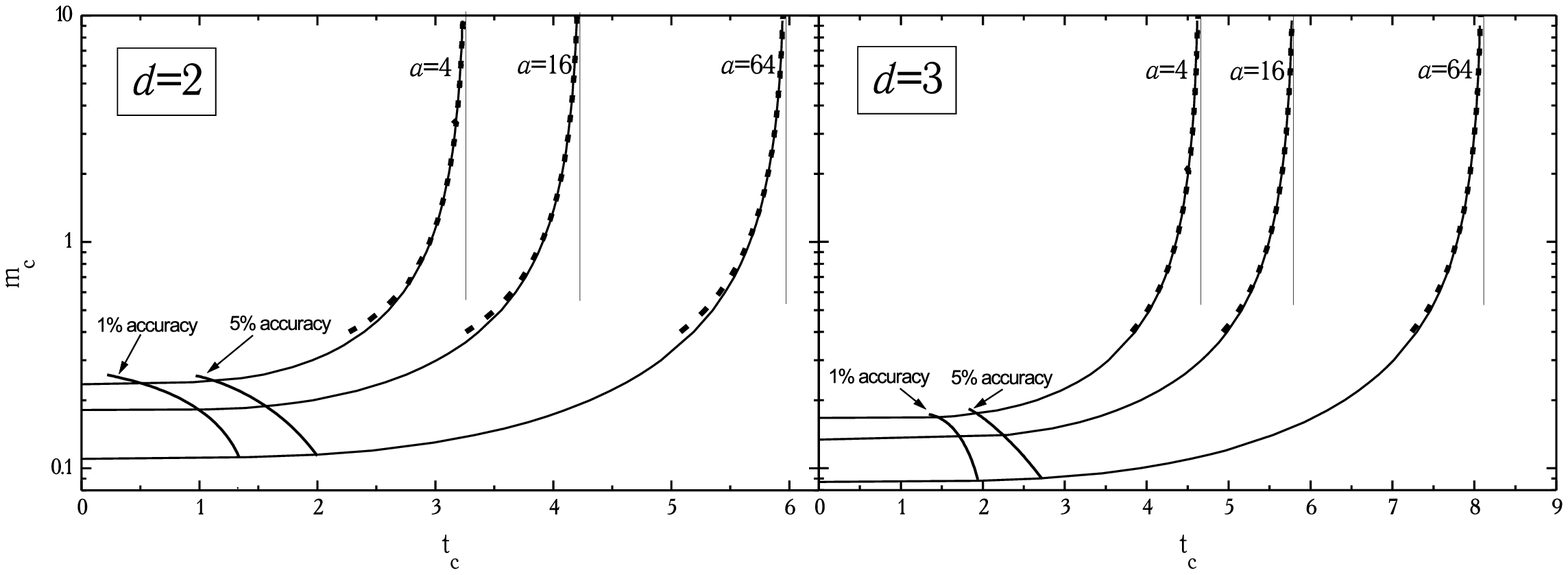}
  \caption{The dependencies of the critical mass $m_c$ on
the critical temperature $t_c$ for the three values of the
parameter $a$: $a=4$, $a=16$, $a=64$. The logarithmic scale for
$m_c$ is used. The left and the right panels show $m_c(t_c)$ for
2D and 3D, respectively. The solid lines show the result of
quantum MF approach ({\ref{mf}}), the vertical lines are the
classical MF limits \cite{AR,my} the dashed lines are the first
quantum correction ({\ref{qcorr}}). The lines marked as 1\% and
5\% show the estimations for the domains, where the "quantum
limit" value of $m_c(t=0)$ delivers the corresponding accuracy for
$m_c(t)$.}
\end{figure}

\begin{figure}[p]
  \centering
  \includegraphics[width=1.0\textwidth]{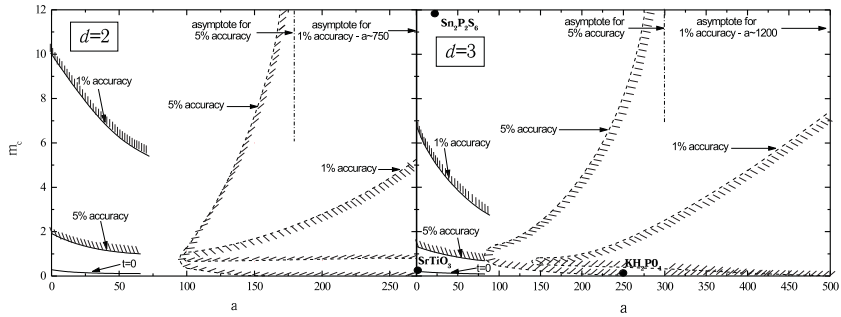}
  \caption{The phase diagram for the limit domains of 2D
and 3D systems (the left and the right panel, respectively). Solid
lines indicate the classical limit $m\to +\infty$ (hatching on
these curves indicates the beginning of the classical area with an
accuracy of 1\% and 5\%). The purely quantum limit corresponds to
the curve with $t=0$ (the domain of applicability for this limit
is shown in Fig.2). The dashed lines indicate the order-disorder
limit area. Vertical lines are the classical asymptotes for these
curves. The displacive limit approximately corresponds to $a<1$.
Points on the right panel are the estimations for $m$ and $a$ from
the known experimental data for Sn$_2$P$_2$S$_6$, SrTiO$_3$, and
KH$_2$PO$_4$.}
\end{figure}

\end{document}